\documentclass[11pt]{article}

\usepackage[T1]{fontenc}
\usepackage[margin=1in]{geometry}
\usepackage{amsmath,amssymb,bm}
\usepackage{graphicx}
\usepackage{booktabs,tabularx,array}
\usepackage{float}
\usepackage[hidelinks]{hyperref}

\def\BibTeX{{\rm B\kern-.05em{\sc i\kern-.025em b}\kern-.08em
    T\kern-.1667em\lower.7ex\hbox{E}\kern-.125emX}}

\newcommand{\Rep}{\mathcal{R}}
\newcommand{\Ops}{\mathcal{A}}
\newcommand{\Compat}{\mathcal{C}}
\newcommand{\Obs}{\mathbf{z}}
\newcommand{\Feat}{\boldsymbol{\phi}}
\newcommand{\Param}{\boldsymbol{\theta}}

\title{Packet-Level In-Network Semantic Adaptation for Unstable Mobile Emergency Networks}
\author{Zhiyuan Ren\thanks{Corresponding author: zyren@xidian.edu.cn}
\qquad Tao Zhang\qquad Wenchi Cheng\\[0.5em]
\small School of Telecommunications Engineering, Xidian University\\
\small Xi'an 710071, China\\
\small zyren@xidian.edu.cn; zhangtao02@xidian.edu.cn; wccheng@xidian.edu.cn}
\date{September 2026}

\begin{document}
\maketitle

\begin{abstract}
Mobile emergency networks can experience independently changing
intermediate wireless links on timescales shorter than endpoint feedback can
track. When an egress changes after packet emission, feedback affects only
later source data, while the on-path node observes the current condition with
the affected packet still mutable. This paper presents DINA, a packet-level
in-network semantic adaptation method. An image is divided into
self-describing spatial packets carrying coordinates, a current representation
identifier, and payload. At each eligible node, an offline-trained frozen
selector scores compatible operators, immediately transforms the packet, and
forwards it without image reconstruction or cross-packet adaptation state.
Later nodes can retain or further compact the packet through the same typed
compatibility contract. The receiver places available packets by coordinate,
fills missing regions with black, and runs a fixed machine task. We realize
DINA in a 24-node UAV environment using XDP and AF\_XDP. In the primary
forest-fire trace, DINA raises deadline tile coverage from 40.4\% to 72.6\%
and classification accuracy from 77.5\% to 95.0\% relative to forwarding. In
an independently trained RescueNet segmentation case, it raises coverage from
65.6\% to 91.7\% and foreground mIoU from 0.486 to 0.541. Sufficient- and
extreme-capacity profiles expose a no-gain boundary and a common task-failure
boundary, respectively.
\end{abstract}


\section{Introduction}
\label{sec:introduction}

Mobile emergency networks are assembled when fixed
infrastructure is unavailable, damaged, or unable to cover a hazardous area.
UAVs, vehicles, and robots act as sensors and temporary relays, carrying
task-bearing observations toward a command endpoint.  Mobility, obstruction,
contention, and changing relay geometry make multi-hop UAV links time-varying
\cite{yao2022uavfleet,hayat2016uav}, while forest measurements show strong
environment-specific propagation effects \cite{yuan2025forestchannel}.
Different directed links on a temporary route can therefore evolve on
different timescales.  A packet can leave its source under one condition and
reach an intermediate node whose outgoing link has since become the path's
immediate constraint.

\begin{figure}[H]
\centering
\includegraphics[width=\linewidth]{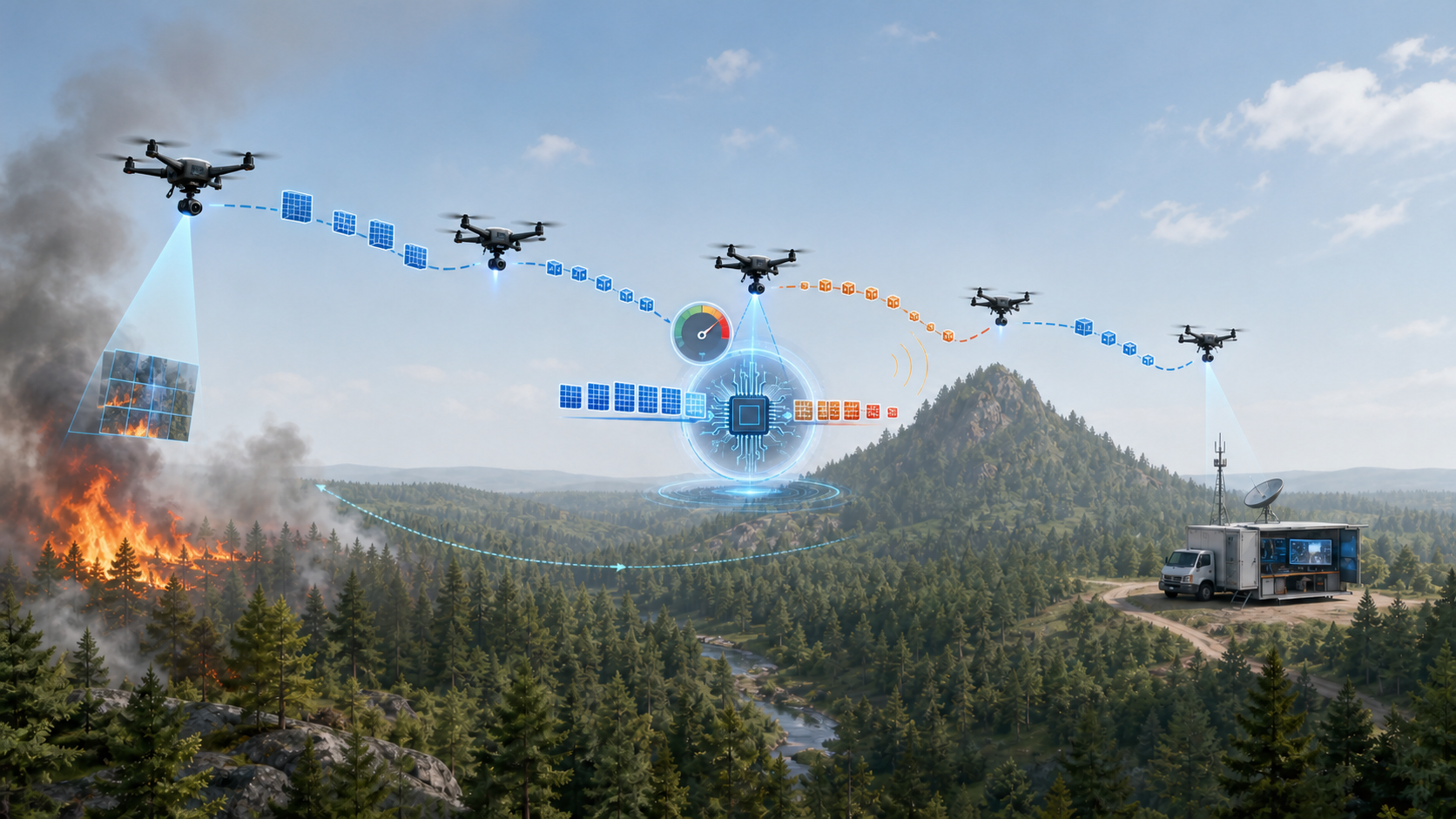}
\caption{DINA in a mobile forest-fire emergency network.  A sensing UAV emits
large RGB spatial packets (blue) through temporary relays.  The highlighted
on-path node observes degradation on its exact outgoing link while those
packets remain mutable, transforms them into smaller task-related packets
(orange/red), and forwards them across the constrained hop.  The faint return
path denotes endpoint feedback.  Under rapid link variation, its observation
can become stale before source-side adaptation takes effect, and it can change
only subsequent packets, not the in-flight packet already present at the
constrained relay.  The evaluated topology and link processes are specified
independently.}
\label{fig:motivating-scenario}
\end{figure}

Endpoint and analytics-edge systems adapt encoder or application settings to
measured resource conditions and task utility
\cite{zhang2018awstream,du2022accmpeg,chen2024regionfilter}.  When the relevant
condition belongs to an intermediate egress, applying that control at the
source requires an observe--return--act loop.  Under rapid variation, the
condition can change again before the source's adapted traffic reaches that
hop.  The action also applies only to data still held by the source and cannot
retroactively change a packet already waiting at the newly constrained relay.
Processing after the weak hop is too late to reduce the load already offered
to it.  As
Fig.~\ref{fig:motivating-scenario} shows, the relay is the first execution point
where the current state of that egress and the affected, still-mutable packet
coexist.

This paper presents \emph{DINA}, a packet-level method that uses this
node-local execution opportunity.  The source divides an image into
self-describing spatial packets carrying coordinates, a current
representation identifier, and its payload.  At each capable on-path node, an
offline-trained and frozen lightweight policy observes the current packet and
the node's exact-egress condition, selects an application-registered operator
compatible with the packet's current representation, transforms the packet,
and forwards it immediately.  The next packet starts from its own inputs; no
image reconstruction or cross-packet adaptation state is involved.

Current-representation typing also supports continued processing along a
multi-hop path.  A downstream node consumes the representation currently
carried by the packet and selects only operators that accept that type.  The
receiver decodes available packets by coordinate, fills missing regions with
black, and runs a fixed machine task on the realized canvas.  Semantic benefit
is therefore measured by the receiver's task outcome, alongside separate
network measurements of delivered coordinates, bytes, queues, and drops.

We realize DINA in a configurable 24-node UAV environment using XDP and
AF\_XDP.  The online worker reads a versioned exact-egress observation,
evaluates a quantized integer selector, and applies a bounded packet-local
operator before returning the packet to ordinary forwarding.  The evaluation
records 41,452 decisions, including 17,452 packets processed at multiple
nodes, with no incompatible transition or relay image reconstruction.  On a
fluctuating forest-fire trace, DINA raises deadline tile coverage from 40.4\%
to 72.6\% and receiver accuracy from 77.5\% to 95.0\% relative to unchanged
forwarding.  A RescueNet segmentation case raises coverage from 65.6\% to
91.7\% and dataset-level foreground mIoU from 0.486 to 0.541 on a composite
trace.  Sufficient-capacity and extreme-capacity profiles locate the measured
no-gain and common task-failure boundaries.

The contributions are as follows:
\begin{itemize}
    \item We identify and formulate the on-path execution opportunity created
    when fresh exact-egress state and a still-mutable in-flight packet coexist
    after that packet has left its source.
    \item We design a self-describing packet and compatible-operator contract
    that supports independent processing and continued representation
    evolution across multiple nodes without image reconstruction or
    cross-packet adaptation state.
    \item We realize the method in an XDP/AF\_XDP packet path with an
    offline-trained, integer-scored selector and evaluate its network and
    receiver-task effects in a 24-node UAV environment across forest-fire
    classification and RescueNet segmentation workloads.
\end{itemize}

\section{Related Work and Positioning}
\label{sec:related}

DINA combines task-aware representation adaptation with execution at the node
that is about to transmit the packet.  Table~\ref{tab:related-position}
compares adjacent approaches by control position, online input, and adaptation
object.

\begin{table}[H]
\centering
\caption{Representative approaches by control position and adaptation object.}
\label{tab:related-position}
\scriptsize
\setlength{\tabcolsep}{2.5pt}
\begin{tabularx}{\linewidth}{>{\raggedright\arraybackslash}p{0.18\linewidth}
>{\raggedright\arraybackslash}p{0.18\linewidth}
>{\raggedright\arraybackslash}p{0.25\linewidth}
>{\raggedright\arraybackslash}X}
\toprule
Line of work & Decision position & Principal online input & Adaptation object \\
\midrule
UAV emergency communication~\cite{bekmezci2013fanet,hayat2016uav,yao2022uavfleet,yuan2025forestchannel}
& Routing, radio, or endpoints & Topology, propagation, channel, and resource state
& Route, coverage, channel use, or endpoint code \\
\addlinespace
ABR, layered coding, and proxy adaptation~\cite{schwarz2007svc,stockhammer2011dash,mao2017pensieve,dasari2022swift,tuker2024packet}
& Server, client, or edge proxy & Throughput, playback buffer, compute, or edge-link rate
& Segment bitrate, coded layer, or removable packet chunk \\
\addlinespace
Task-aware visual analytics~\cite{zhang2017videostorm,jiang2018chameleon,zhang2018awstream,li2020reducto,du2022accmpeg,chen2024regionfilter}
& Camera, source, or analytics edge & Content profile, task accuracy, bandwidth, and compute
& Frame selection, resolution, model configuration, or encoding allocation \\
\addlinespace
Semantic communication and split inference~\cite{bourtsoulatze2019deepjscc,xie2021semantic,strinati2021semantic,uysal2022semantic,kang2017neurosurgeon,shao2020bottlenet}
& Learned endpoints or a DNN partition & Channel state, task loss, and model state
& Channel symbols, latent features, or partition point \\
\addlinespace
Programmable in-network computing~\cite{sapio2017dumb,liu2017incbricks,jin2017netcache,jin2018netchain,gupta2018sonata,sapio2021switchml,zheng2023dinc}
& Switch, NIC, kernel, or attached worker & Application state and data-plane resources
& Cached values, aggregates, queries, or packets \\
\addlinespace
Network-side content adaptation~\cite{chen2023octopus,tuker2024packet}
& Base station or edge function & Local capacity, queue state, and content priority
& Marked message, enhancement layer, or packet chunk \\
\addlinespace
DINA & Capable node before its current egress & Current packet, representation, and node-local observation
& Current spatial-packet payload and representation type \\
\bottomrule
\end{tabularx}
\end{table}

\subsection{UAV Emergency Networks}

Flying ad hoc networks and civil UAV systems use airborne nodes as sensors,
relays, and temporary infrastructure under mobility and topology change
\cite{bekmezci2013fanet,hayat2016uav}.  Emergency-fleet, airborne-coverage,
and forest-channel studies further characterize harsh deployment conditions
\cite{yao2022uavfleet,alhourani2014lap,yuan2025forestchannel}.  This literature
establishes DINA's dynamic multi-hop setting.  DINA takes the route supplied by
the underlying network and acts on the current packet when its next-hop
condition changes; its adaptation object is the packet representation rather
than the topology, radio resource, or route.

\subsection{Endpoint and Task-Aware Adaptation}

Adaptive streaming changes data controlled at an endpoint: scalable coding
and DASH expose layers or segments, while Pensieve and Swift learn policies or
use layered neural codecs
\cite{schwarz2007svc,stockhammer2011dash,mao2017pensieve,dasari2022swift}.
Task-aware visual analytics adapts model configuration, frame selection,
resolution, encoding, and region filtering according to resource and receiver
utility~\cite{zhang2017videostorm,jiang2018chameleon,zhang2018awstream,
li2020reducto,du2022accmpeg,chen2024regionfilter}.  These systems establish
task-guided reduction at the source, client, or analytics edge.

Learned semantic communication instead optimizes transmitted symbols or
features for reconstruction or task goals
\cite{bourtsoulatze2019deepjscc,xie2021semantic,strinati2021semantic,
uysal2022semantic}, including digital--analog transmission for emergency
communication~\cite{fu2025daesemcom}.  Collaborative inference partitions a
DNN or compresses an intermediate feature between device and edge
\cite{kang2017neurosurgeon,shao2020bottlenet}.  DINA lets the receiver task
define registered representations, while only a frozen choice and bounded
packet transform run on path.  The complete task model remains at the
receiver, and the local action applies to the current in-flight packet.

\subsection{Network-Side Reduction}

Edge packet trimming removes less important chunks from layered video using
edge capacity \cite{tuker2024packet}.  Octopus uses actual 5G base-station
capacity to discard application-marked messages or enhancement layers
\cite{chen2023octopus}.  Both demonstrate that a network-side execution point
can react to local service conditions before traffic enters a constrained
access link.  Recent emergency-network work also places light-model switching
inside the network~\cite{li2026boundswitch}, showing a complementary form of
task-related local execution.

DINA builds on this evidence for network-side actuation but changes a
different object.  Rather than only dropping a marked unit, removing an
enhancement chunk, or changing an inference model, a DINA node converts the
payload of the current spatial packet into an application-registered
representation.  The packet records its resulting type, so a later node can
select another compatible transform from the representation actually
received.  Typed continuation therefore lets one packet's representation
evolve across independently changing hops.

\subsection{Programmable In-Network Computing}

Programmable data planes execute caching, coordination, telemetry, and
distributed-training aggregation in switches or attached processors
\cite{sapio2017dumb,liu2017incbricks,jin2017netcache,jin2018netchain,
gupta2018sonata,sapio2021switchml}.  P4, XDP, and AF\_XDP provide practical
substrates for bounded on-path execution
\cite{bosshart2014p4,hoiland2018xdp,linux2026afxdp}; recent systems distribute
functions across devices, run learned traffic analysis in a switch, and
support payload-mutating functions
\cite{zheng2023dinc,yan2024brain,ji2025mtp}.  DINA supplies an
application--network contract for this substrate: a self-describing packet,
typed transforms, and a node-local representation policy connect packet
processing to a receiver-side machine task.

\section{Scenario and Problem Formulation}
\label{sec:model}

\subsection{Unstable Multi-Hop Image Return}

Let the mobile emergency network at time $t$ be a directed graph
$G(t)=(V,E(t))$.  A source $s\in V$ sends an image object $I_f$ to destination
$d\in V$ over the route
\begin{equation}
P_f(t)=(v_0=s,v_1,\ldots,v_{H_f}=d).
\end{equation}
The underlying network supplies the route and may change it with topology.
For directed link $e=(v,u)$, node $v$ reads a local observation
\begin{equation}
\Obs_e(t)=\big(c_e(t),q_e(t),d_e(t),\ell_e(t),\ldots\big),
\end{equation}
where the components can include current service rate, queued bytes, recent
delivery counters, connectivity, or other locally exported measurements.  The
observation is local to $v$ and is sampled when the packet is processed.

\subsection{Where an In-Network Action Can Help}

For a packet waiting at node $v$, changing its representation can affect only
links that the packet has not crossed.  Let $e=(v,u)$ be its physical next
hop.  A useful execution opportunity exists when two conditions hold at the
same location and time:
\begin{equation}
\begin{aligned}
F_e(t)&=\mathbb{I}\{\operatorname{age}(\Obs_e(t))\leq\tau_e\},\\
M(p,e,t)&=\mathbb{I}\{\mathsf{mutable}(p,e,t)\},\\
\mathcal{O}(p,v,e,t)&=F_e(t)M(p,e,t).
\end{aligned}
\label{eq:opportunity}
\end{equation}
The first term requires a sufficiently fresh application observation.
$\mathsf{mutable}(p,e,t)$ means that the node can still change the packet
before it is committed to egress service.  Any on-path node satisfying both
terms can execute DINA.

This location matters because different hops in a mobile route need not
degrade together.  End-to-end feedback can describe a path-level consequence,
but its observation must return to the source and source-adapted packets must
then travel toward the affected hop.  If that egress changes within this loop,
the source decision may no longer match its condition when those packets
arrive.  Feedback also cannot modify a packet that has already left the source,
and processing after the constrained link cannot reduce the load already
offered to it.  DINA therefore treats the route as a sequence of local
execution opportunities, each defined by the physical egress that the packet
is about to traverse.  The evaluation measures both network delivery and the
receiver task because a smaller payload can still discard useful task evidence.

\subsection{Spatial Packet Object}

Each $480\times320$ image in the current case is divided into $N=600$
$16\times16$ spatial packets.  Packet $p_{f,i}$ contains
\begin{equation}
p_{f,i}=\langle f,x_i,y_i,w_i,h_i,r_{f,i},b_{f,i}\rangle,
\label{eq:packet}
\end{equation}
where $f$ is the image identifier, $(x_i,y_i,w_i,h_i)$ is placement geometry,
$r_{f,i}\in\Rep$ is the current representation, and $b_{f,i}$ is the current
payload.  Sequence position is not needed for node processing or receiver
placement.

The packet object separates application knowledge from network execution.  The
application defines how a spatial region is represented and decoded; the
network receives a typed, bounded object that can be transformed without the
complete image.  Workloads with cross-packet dependencies require an
application packet contract that exposes an independently processable object.

\subsection{Packet-Local Semantic Adaptation}

The application registers a finite set of operators $\Ops$.  Each operator
$a\in\Ops$ declares an input type, output type, and packet-local transform
\begin{equation}
a:\mathcal{X}_{r_{\mathrm{in}}}\rightarrow
  \mathcal{X}_{r_{\mathrm{out}}}.
\end{equation}
For current representation $r$, the immutable compatibility registry returns
\begin{equation}
\Compat(r)=\{a\in\Ops:\,r\in\operatorname{input}(a)\}.
\end{equation}
When $p_{f,i}$ reaches node $v$ at time $t_{f,i}^{v}$, DINA chooses
\begin{equation}
a_{f,i}^{v}=
\pi_{\Param}\!\left(p_{f,i}^{v},\Obs_{(v,u)}(t_{f,i}^{v}),
\Compat(r_{f,i}^{v})\right),
\label{eq:selector}
\end{equation}
where $u$ is the packet's next hop and $\Param$ is trained offline and frozen
during a run.  The packet is then updated as
\begin{align}
b_{f,i}^{v+}&=a_{f,i}^{v}(b_{f,i}^{v}),\\
r_{f,i}^{v+}&=T(r_{f,i}^{v},a_{f,i}^{v}),
\end{align}
and forwarded.  Retention is an explicit action that leaves the current
representation unchanged.

The online state used by Eq.~\eqref{eq:selector} ends with the packet.  Two
packets of the same image can select different operators when they encounter
different node-local observations, and coordinate-based placement makes their
delivery order irrelevant.

For a single packet, the current representation can change at several nodes.
Suppose node $v_j$ converts RGB24 to FIRE8 and a later node $v_{j+2}$ receives
that packet.  The later node filters its choices through
$\Compat(\mathrm{FIRE8})$ and may retain FIRE8 or produce a representation
whose operator accepts FIRE8.  It cannot schedule an RGB-only transform or
recreate discarded color channels.  Only the current type and payload are
needed to determine the next compatible action.

If a link recovers, a later RGB24 packet from the same image can remain RGB24
even though an earlier packet was compacted.  A subsequent degradation can
produce another representation.  The realized image therefore follows packet
arrivals and local observations without an image-level transition rule.

\subsection{Receiver Task and Performance Boundary}

At a configured source-relative time $D_f$, the destination initializes a
black canvas $\widehat I_f(D_f)$ and overwrites every coordinate for which a
packet is available.  The representation identifier determines how the payload
is decoded into that region.  A fixed receiver model $h$ then produces
\begin{equation}
\widehat y_f=h(\widehat I_f(D_f)).
\end{equation}
Semantic performance is measured against task label $y_f$ through the fixed
receiver model.

The receiver runs the task at the configured evaluation time and records the
coordinates and representations present on the canvas when $h$ is invoked.

For a fixed image stream, deadline, topology process, and operator library,
the operating space contains three empirically distinguishable regions:
\begin{itemize}
    \item a \emph{sufficient} region in which unchanged RGB forwarding already
    supports the task;
    \item an \emph{intermediate} region in which compact packet
    representations can change delivery and receiver-task outcomes; and
    \item an \emph{extreme} region in which even the most compact registered
    representation leaves insufficient task evidence.
\end{itemize}
Their measured boundaries depend on the workload, deadline, queue, registry,
and receiver.

\section{DINA Design}
\label{sec:design}

\subsection{Architecture}

\begin{figure}[H]
\centering
\includegraphics[width=\linewidth]{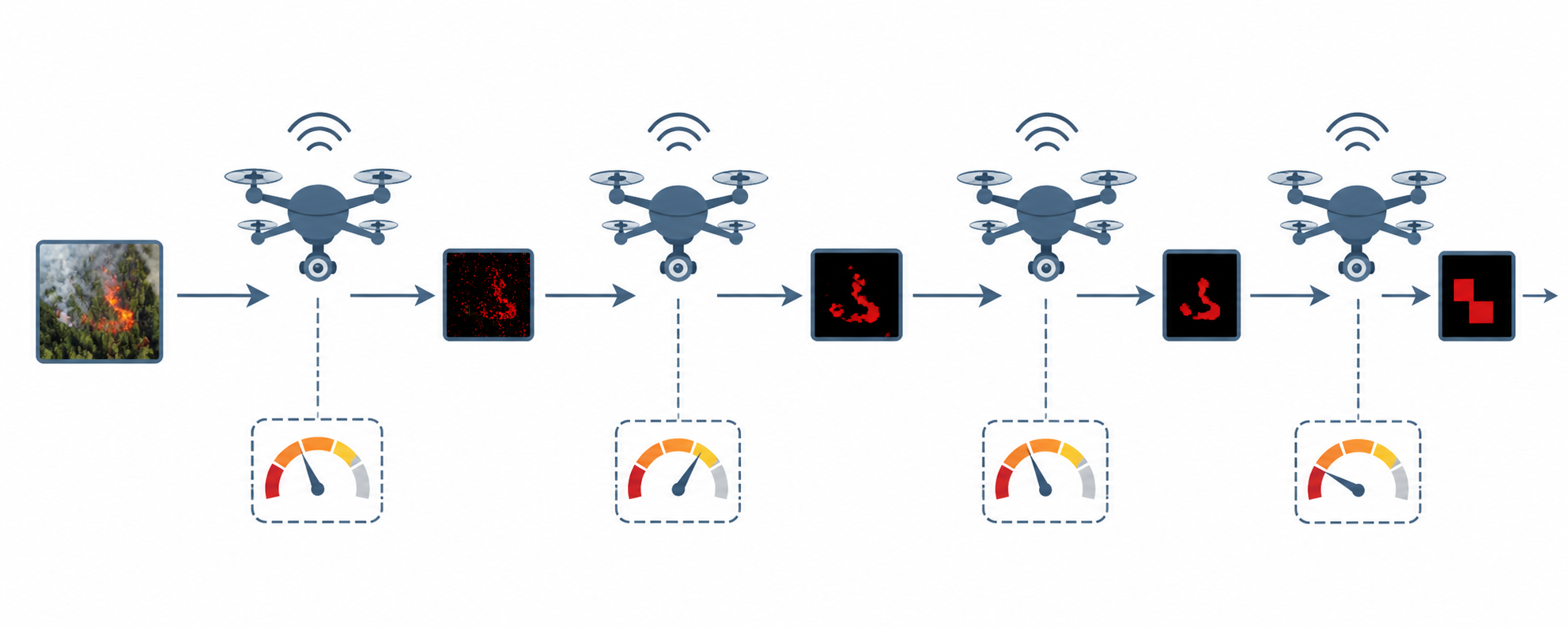}
\caption{Packet-local representation evolution along a multi-hop path.  Each
node observes its own outgoing condition, processes only the current packet,
updates the packet's current representation, and returns it to forwarding.
No intermediate node reconstructs the image.}
\label{fig:architecture}
\end{figure}

Fig.~\ref{fig:architecture} shows the execution boundary.  The source performs
spatial packetization once.  Routing forwards each packet through any number of
eligible nodes.  At a node, XDP identifies DINA traffic and redirects it to the
local AF\_XDP worker.  The worker reads the exact outgoing-link record, selects
and executes an operator, and forwards the updated packet.  Full machine
inference occurs only at the destination.

The ordinary network plane supplies topology, route selection, queueing, and
link service.  The DINA packet plane performs parsing, compatibility filtering,
operator selection, transformation, and representation update.  The
application supplies operator definitions, offline training material, the
coordinate decoder, and the receiver task.

A packet can encounter zero, one, or several eligible nodes.  Enabled nodes
return it to the route supplied by ordinary forwarding after bounded
processing.  Resolving the next hop before selection binds the worker's
observation to the exact physical egress.

\subsection{Self-Describing Packet Contract}

Table~\ref{tab:packet-contract} lists the logical packet fields.  Object and
geometry fields let the receiver place a packet independently.  The current
representation makes both downstream interpretation and compatibility checks
independent of operation history.  Wire length is updated after a transform so
the compact representation reduces the load offered to the following link.

Let $H_p$ denote the fixed packet and transport overhead and let
$L_a(w,h)$ be the serialized payload produced by representation $a$ for a
$w\times h$ region.  The load offered to the next hop after processing is
\begin{equation}
B(p,a)=H_p+L_a(w,h).
\label{eq:wire-load}
\end{equation}
For a set $\mathcal P_e(\Delta)$ of packets presented to egress $e$ during
interval $\Delta$, the transformed offered load is
\begin{equation}
B_e(\Delta)=\sum_{p\in\mathcal P_e(\Delta)}B(p,a_p).
\label{eq:egress-load}
\end{equation}
These equations account for actual serialized bytes; an operator is not
credited with a networking benefit merely because its decoded canvas looks
sparse.  Headers, metadata, and any packet that is retained remain in the wire
budget.

\begin{table}[t]
\centering
\caption{Logical packet contract.}
\label{tab:packet-contract}
\begin{tabular}{p{0.23\columnwidth}p{0.65\columnwidth}}
\toprule
Field & Purpose \\
\midrule
Object identifier & Associates independently arriving tiles with one image. \\
Coordinates and geometry & Determines placement without arrival-order assumptions. \\
Current representation & Selects payload decoder and compatible downstream actions. \\
Payload length & Describes the current wire payload after transformation. \\
Current payload & Carries the original or task-related spatial representation. \\
\bottomrule
\end{tabular}
\end{table}

\subsection{Task-Related Operator Registry}

Operators are supplied and profiled by the application.  The forest-fire case
instantiates the interface with the simple chain in
Table~\ref{tab:operators}.  A shared color rule marks fire-candidate
pixels.  FIRE8 retains the red intensity at those positions,
FIRE1 retains the binary candidate mask, and
FIRE1-DS4 applies $4\times4$ max pooling to that mask.  These operations
execute on one tile without neighboring tiles or the receiver model.

\begin{table}[t]
\centering
\caption{Registered representations for a $16\times16$ tile.}
\label{tab:operators}
\small
\setlength{\tabcolsep}{3pt}
\begin{tabular}{p{0.19\columnwidth}p{0.13\columnwidth}p{0.27\columnwidth}p{0.28\columnwidth}}
\toprule
Rep. & Payload & Accepted inputs & Meaning \\
\midrule
RGB24 & 768 B & RGB24 & retain RGB pixels \\
FIRE8 & 256 B & RGB24, FIRE8 & candidate red intensity \\
FIRE1 & 32 B & RGB24, FIRE8, FIRE1 & candidate mask \\
FIRE1-DS4 & 2 B & all registered types & pooled mask \\
\bottomrule
\end{tabular}
\end{table}

\begin{figure}[H]
\centering
\includegraphics[width=\linewidth]{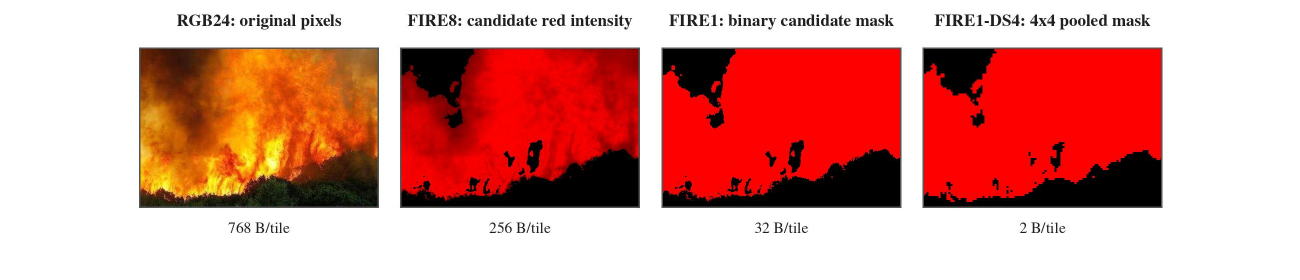}
\caption{Full-scene outputs of the registered forest-fire case-study
representations on the same held-out source image.  Every panel is produced by
applying the current packet-local operator independently to $16\times16$
tiles and reconstructing them by coordinate.  From left to right, the four
panels show progressively smaller serialized tile payloads; byte counts exclude
common headers.}
\label{fig:operator-outputs}
\end{figure}

Figure~\ref{fig:operator-outputs} makes the application contract concrete.
RGB24 preserves the complete scene.  FIRE8 preserves one byte of candidate
red intensity per pixel.  FIRE1 keeps only the candidate predicate, packed as
one bit per pixel.  FIRE1-DS4 pools each $4\times4$ group inside the tile and
therefore carries 16 bits.  The receiver expands each representation to its
spatial region before running the fixed task model.  No intermediate node
needs to know whether the scene contains fire; it only executes a registered
transform selected by the lightweight policy.

Compatibility is encoded as a directed representation graph.  A packet can
follow
\begin{equation}
\mathrm{RGB24}\rightarrow\mathrm{FIRE8}\rightarrow
\mathrm{FIRE1}\rightarrow\mathrm{FIRE1\_DS4},
\end{equation}
or skip to any registered output that accepts its current input.  Retention at
a downstream node is always possible.  A transform toward a representation
requiring information absent from the current payload is rejected before
scoring.

The graph is immutable during a run and checked independently at every node.
The learned policy can retain a packet at one node and compact it at another;
a later packet from the same image can make a different choice.  A transition
is executable when its declared input accepts the current payload.

\subsection{Packet Processing Procedure}

The critical runtime is summarized in Fig.~\ref{fig:packet-procedure}.  The
deployed selector evaluates frozen learned scores after compatibility
filtering.

\begin{figure}[t]
\centering
\fbox{\begin{minipage}{0.92\columnwidth}
\small
\textbf{Packet-local node procedure}\par
\vspace{2pt}
\textbf{Input:} packet $p$, physical egress $e$, frozen registry and selector.\par
1. Parse coordinates, current representation $r$, and payload.\par
2. Read one consistent current observation $\Obs_e(t)$.\par
3. Construct $\Compat(r)$ from immutable type metadata.\par
4. Score each $a\in\Compat(r)$ using the frozen selector.\par
5. Select the highest-scoring action; resolve an exact tie in favor of the
more informative compatible representation.\par
6. Apply the action to this payload, update $r$ and wire length, and forward.\par
7. Discard packet-local variables when the packet leaves the worker.
\end{minipage}}
\caption{The DINA online procedure.  Each invocation completes for one packet.}
\label{fig:packet-procedure}
\end{figure}

The procedure is deliberately restartable for every packet.  All mutable
variables---the parsed header view, feature vector, compatible-action mask,
scores, and transformed length---belong to the current invocation.  Immutable
objects such as the operator registry and frozen selector weights can be shared
by all packets without creating adaptation history.  Queue state belongs to
the network observation and may change between invocations; DINA samples it as
an input to each packet decision.

Representation retention is included in the action set.  Exact ties resolve
toward the more informative compatible representation, preventing quantized
ties from triggering arbitrary compaction.  Parser failures, unknown
representations, and invalid transforms are counted as implementation errors.

\section{Offline Training and Online Selection}
\label{sec:selector}

DINA uses an offline-trained policy that is frozen for bounded online
evaluation.  For training item $q$, $\Feat_q$ contains packet-visible features
and network observations available at an execution node, and $U_{q,a}$ is the
counterfactual utility of compatible action $a$.  Image and trace groups are
disjoint across training, validation, and test material.  At runtime,
\begin{equation}
a^*=\arg\max_{a\in\Compat(r)}Q_{\Param}
\big(a\mid\Feat(p,\Obs_e(t))\big).
\label{eq:abstract-score}
\end{equation}
The feature function reads the current packet and local observation.  The
implemented instance uses connectivity, current-to-design rate ratio,
exact-egress queue pressure, current-representation rank, and payload ratio.
We fit one ridge-regression value function per action and quantize the
coefficients into the signed-integer \texttt{DINA\_SELECTOR\_V1} ABI.  The
AF\_XDP runtime evaluates
\begin{equation}
a^*=\arg\max_{a\in\Compat(r)}
\left\langle\widehat{\Param}_a,\Feat(p,\Obs_e(t))\right\rangle,
\label{eq:online-score}
\end{equation}
where $\widehat{\Param}_a$ is loaded at worker startup.  A parity harness feeds
identical current representations and feature vectors to the Python evaluator
and compiled C selector; accepted exports match on integer features,
compatibility masks, per-action scores, and selected actions.  The weights
remain immutable throughout each run.

\section{Implementation}
\label{sec:implementation}

\subsection{Twenty-Four-Node UAV Environment}

The maintained environment contains 24 homogeneous network-node containers in
a configurable UAV mesh.  A topology manager publishes connectivity and
directed-link parameters, and the existing routing subsystem derives paths
from that topology.  Directed-link profiles change independently and
asynchronously without resetting queues or application objects.  Different
nodes on one active route can therefore observe different service regimes,
and an in-flight packet can encounter a condition that differs from the one at
source emission.  The environment records each realized route and node-local
worker decision.

The primary workload is a continuous stream of independent forest-fire images,
with 600 self-contained packets per image.  Congestion changes which spatial
packets and representations are available when the receiver constructs each
canvas.  The RescueNet case retains the same packet geometry and path while
using a gray/edge registry and a semantic-segmentation receiver.

\subsection{XDP and AF\_XDP Packet Path}

XDP performs early identification and redirection.  An AF\_XDP worker inside
the network-node container owns the DINA execution step and returns the packet
to normal forwarding after processing.  AF\_XDP provides shared packet rings
and user-space access without placing the complete image task in the kernel
\cite{linux2026afxdp}.  The worker contains packet parsers, the compatibility
table, four tile operators, and the frozen integer selector.

The packet path consists of five bounded stages.  XDP classifies an
eligible packet and redirects it to the AF\_XDP socket.  The host forwarding
fabric resolves the next hop and inserts a fixed node-local selector entry
whose argument identifies the physical egress.  The worker parses the current
representation, reads one consistent observation record, and evaluates the
compatible actions.  It then rewrites the payload, representation identifier,
and length if the chosen action transforms the packet.  Finally, the packet is
returned to the ordinary forwarding path and the node-local entry is marked as
visited for that packet.

The visited entry is a packet-local loop guard that prevents immediate
redirection to the same worker.  If forwarding later reaches another eligible
node, that node performs a new decision using its own egress record and the
packet's updated representation.  Learned runs load an explicit selector file
at startup and record its identity in the run manifest.

\subsection{Read-Only Observation Interface}

The host exposes a versioned read-only exact-egress table containing
connectivity, current and design rate, queue occupancy and capacity,
diagnostic counters, sample time, and topology generation.  Each worker maps
the table once and performs no JSON parsing, socket request, or file write in
the packet path.  A sequence lock and schema, generation, and next-hop checks
prevent workers from consuming a partially updated or mismatched link record.
The accepted record identity is retained in the packet event for later audit.

\subsection{Receiver Reconstruction}

The receiver initializes a $480\times320$ black canvas for every object.  It
decodes each packet according to the packet's current representation and
places the resulting tile at the encoded coordinate, independent of arrival
order.  At the configured source-relative evaluation time, the primary
receiver applies one frozen YOLO model and records the task result.  Later
arrivals do not rewrite that result.  Every network method uses the same
receiver model, preprocessing, threshold, coordinate placement, and black-fill
rule; intermediate containers neither contain the receiver checkpoint nor
query its predictions.

The RescueNet receiver uses the same coordinate placement and black-fill
rule.  Its application registry contains RGB24, Gray8, Gray4, and Edge1, and
its frozen DeepLabV3--MobileNetV3 model produces an 11-class segmentation.
Training, validation, and test separation for both receivers is specified in
Section~\ref{sec:evaluation}.  Packet, forwarding, reconstruction, and task
records share one object identifier for end-to-end reconciliation and
reproduction of naturally mixed canvases.

\section{Evaluation Methodology}
\label{sec:evaluation}

The evaluation is organized around the questions in
Table~\ref{tab:research-questions}.  Every reported network comparison uses
paired inputs and a source-relative decision time.  Calibration runs used to
locate operating points are excluded from the reported test results.

\begin{table}[H]
\centering
\caption{Evaluation questions and required evidence.}
\label{tab:research-questions}
\small
\setlength{\tabcolsep}{3pt}
\begin{tabularx}{\linewidth}{>{\raggedright\arraybackslash}p{0.07\linewidth}
>{\raggedright\arraybackslash}X
>{\raggedright\arraybackslash}X}
\toprule
RQ & Question & Evidence \\
\midrule
RQ1 & Can the fixed receiver interpret registered and naturally mixed
representations? & Held-out task metrics for complete representations, packet
mixtures, and black-filled canvases. \\
RQ2 & Does a packet undergo compatible processing at multiple nodes without
image state? & Per-node transition trace, input/output type audit, local-rate
generation audit, and absence of image reconstruction. \\
RQ3 & Where does network reduction translate into a receiver-task change? &
Paired sufficient, intermediate, and extreme profiles for DINA and unchanged
forwarding. \\
RQ4 & Does the mechanism remain responsive under asynchronous link and route
conditions? & Packet representation, queue, coverage, and task outcomes under
a continuously varying multi-hop profile. \\
RQ5 & What does packet-local execution cost? & Measured online-decision
latency, paired worker CPU and memory, and packet metadata. \\
RQ6 & Does the same packet path remain useful for a different emergency
machine task? & Frozen RescueNet segmentation model, task-specific registered
representations, three static capacity regions, and one composite trace. \\
\bottomrule
\end{tabularx}
\end{table}

\subsection{Dataset and Receiver Model}

The forest-fire dataset is divided by source group so that visually derived
variants of one source cannot cross training, validation, and held-out test
partitions.  The receiver is trained on RGB24, FIRE8, FIRE1, FIRE1\_DS4,
packet-wise mixtures, and black-filled missing regions.  Model seed and
decision threshold are selected using validation data only, after which the
same model and threshold are frozen for every network method.

\begin{table}[t]
\centering
\caption{Primary forest-fire workload parameters.}
\label{tab:parameters}
\small
\setlength{\tabcolsep}{3pt}
\begin{tabular}{p{0.41\columnwidth}p{0.47\columnwidth}}
\toprule
Parameter & Frozen value \\
\midrule
Training/validation/test source images & 2,996 / 736 / 737 \\
Network-test images (fire/non-fire) & 40 (20/20) \\
Image resolution and tiles & $480\times320$, 600 tiles \\
Object period / packet pacing & 13 s / 15 ms \\
Source-relative decision time & 12 s \\
Queue size and drop policy & 8 KiB / tail drop \\
Topology and routing & 24 UAVs / dynamic two-hop primary routes \\
Link traces & deterministic, phase-aligned 10-epoch profiles \\
Paired test objects per point & 40 \\
\bottomrule
\end{tabular}
\end{table}

The bounded second case uses the official RescueNet post-disaster UAV
segmentation release~\cite{rahnemoonfar2023rescuenet}.  It contains 3,595
training pairs, 449 validation pairs,
and 450 reserved test pairs, each with an 11-class mask.  Two
DeepLabV3--MobileNetV3 candidates receive the same 40-epoch budget on
deterministic registered mixtures; validation foreground mIoU selects one
checkpoint before any test execution.  The formal network playlist contains
40 test images chosen by a content-hash order that does not read labels, model
outputs, or network results.  Table~\ref{tab:rescuenet-parameters} records the
network contract shared by its 24 method--region runs.

\begin{table}[t]
\centering
\caption{Frozen RescueNet network-test contract.}
\label{tab:rescuenet-parameters}
\small
\setlength{\tabcolsep}{3pt}
\begin{tabular}{p{0.43\columnwidth}p{0.45\columnwidth}}
\toprule
Parameter & Frozen value \\
\midrule
Official train/validation/test pairs & 3,595 / 449 / 450 \\
Network-test images & 40 test images \\
Registered representations & RGB24, Gray8, Gray4, Edge1 \\
Receiver metric & Foreground mIoU \\
Image geometry & $480\times320$, 600 tiles \\
Object period / packet pacing & 13 s / 15 ms \\
Canvas window / queue & 12 s / 8 KiB tail drop \\
Primary / backup route & 24--4--3--7--17 / 24--14--13--18--17 \\
\bottomrule
\end{tabular}
\end{table}

\subsection{Compared Methods}

The primary comparison is between DINA and forwarding-only, which retains each
packet's current representation.  Fixed-representation controls traverse the
same worker and operator path but apply one named representation without live
link features.  Every paired run shares topology, route realization, image
order, pacing, coordinates, link trace, queue, decision time, canvas
construction, and receiver model.  Octopus and packet trimming are compared
at the mechanism level in Section~\ref{sec:related} because their application
contracts expose droppable messages or layers instead of transformable spatial
packets.

\subsection{RQ1: Registered and Mixed Task Inputs}

Held-out images are materialized as every complete representation,
deterministic packet-wise mixtures, and mixtures with black coordinates.
Source-group splits keep all variants of one scene together.  We report the
receiver confusion matrix, accuracy, precision, recall, false-fire rate, F1,
F2, and balanced accuracy; network runs then evaluate their realized canvases
directly.

\subsection{RQ2: Multi-Node Execution Audit}

We trace packets processed by at least two nodes and audit the physical egress,
topology generation, observation, input/output representation, compatibility
mask, and wire length at every decision.  A run is valid when transitions are
compatible, events reconcile, no relay reconstructs an image, and coordinate
placement succeeds under reordered arrival.

\subsection{RQ3: Paired Performance Boundary}

Validation-only calibration locates sufficient, intermediate, and extreme
operating regions.  The profiles are then frozen and evaluated on paired
held-out images at a source-relative 12-second decision time.  Their rates are
interpreted together with the configured offered load, packet sizes, queue,
and receiver task.

\subsection{RQ4: Unstable Links}

The intermediate profile changes local links throughout packet emission and
can trigger route changes.  Each worker decision is joined to its exact-egress
rate and queue observation on a common source-relative timeline, together
with the packet's input and selected representation.

\subsection{RQ5: Packet-Path Cost}

The worker times exact-egress observation, feature construction, selector
inference, an optional transform, and checksum update after parsing and before
forwarding.  Thirty-second container samples provide paired CPU and memory
measurements; header sizes come from the compiled packet ABI.

\subsection{RQ6: Bounded Cross-Task Reuse}

The RescueNet case changes the application registry and receiver task while
retaining the packet ABI, coordinate reconstruction, black fill, egress
observation, compatibility filtering, selector interface, and XDP/AF\_XDP
path.  DINA, forwarding, and four fixed representations share a 40-image
playlist, frozen routes, and sufficient, intermediate, extreme, and composite
profiles.

\subsection{Metrics and Paired Analysis}

Network metrics are tile coverage and payload bytes at the source-relative
time, congestion drops, queue occupancy, and token or queue waiting.  Task
metrics are accuracy, fire precision, fire recall, false-fire rate, F1,
balanced accuracy, and the confusion matrix.  RescueNet reports dataset-level
foreground mIoU from the accumulated pixel confusion matrix and the paired
per-image mIoU distribution.

For image $f$, deadline tile coverage is
\begin{equation}
\operatorname{cov}_f(D_f)=
\frac{\left|\{i:p_{f,i}\text{ is placed by }D_f\}\right|}{N}.
\label{eq:coverage}
\end{equation}
Delivered bytes are counted at the receiver by the same time and reported
separately from coverage.  Paired image identifiers and identical traces
support exact sign tests for coverage and McNemar tests for classification
correctness.  RescueNet foreground mIoU is accumulated over all 40 pixel
confusion matrices; paired per-image differences use 10,000 fixed-seed
percentile-bootstrap resamples, with undefined pairs reported explicitly.

\section{Evaluation Results}
\label{sec:results}

\subsection{Receiver Capability across Registered Inputs}

Table~\ref{tab:receiver-capability} evaluates the frozen receiver on 737
source-group-disjoint test images.  Each complete representation contributes
one canvas per source image.  The packet-wise rows aggregate four
deterministic mixtures each, either without missing tiles or with 10\% and
25\% black-filled coordinates.  All six input families retain useful fire
evidence under the same validation-selected decision threshold.  In
particular, naturally mixed canvases achieve 95.66\% accuracy and 97.21\%
fire recall; adding the evaluated black-filled masks changes these values only
to 95.56\% and 96.68\%, respectively.  These measurements establish the
receiver input family used by the case study; each realized network canvas is
still evaluated directly.

\begin{table}[H]
\centering
\caption{Receiver capability on source-group-disjoint test inputs.}
\label{tab:receiver-capability}
\scriptsize
\setlength{\tabcolsep}{2.5pt}
\begin{tabularx}{\linewidth}{>{\raggedright\arraybackslash}Xrrrrrr}
\toprule
Input family & Samples & Accuracy & Precision & \shortstack{Fire\\recall} & F1 &
\shortstack{False-fire\\rate} \\
\midrule
RGB24 & 737 & 0.9213 & 0.8564 & 0.9909 & 0.9188 & 0.1355 \\
FIRE8 & 737 & 0.9471 & 0.9101 & 0.9789 & 0.9432 & 0.0788 \\
FIRE1 & 737 & 0.9376 & 0.8926 & 0.9789 & 0.9337 & 0.0961 \\
FIRE1-DS4 & 737 & 0.9362 & 0.8923 & 0.9758 & 0.9322 & 0.0961 \\
Packet-wise mixtures & 2,948 & 0.9566 & 0.9340 & 0.9721 & 0.9526 & 0.0560 \\
Mixtures with black fill & 2,948 & 0.9556 & 0.9364 & 0.9668 & 0.9513 & 0.0536 \\
\bottomrule
\end{tabularx}
\end{table}

\subsection{Packet-Local Execution across Multiple Nodes}

The execution audit follows 24,000 source packets through a continuously
changing 24-node topology.  The active primary route alternates between
$24\!\rightarrow\!18\!\rightarrow\!17$ and
$24\!\rightarrow\!23\!\rightarrow\!17$.  Of the source packets, 17,452 are
examined by workers at two or more on-path nodes.  The trace records 41,452
packet-local decisions, including 24,006 actual representation changes and
17,446 compatible unchanged outcomes.  Later workers consume the representation
carried by each packet; no operation-history or image-progress state appears
in the worker decision record.

\begin{table}[t]
\centering
\caption{Multi-node packet-execution audit.}
\label{tab:execution-audit}
\small
\setlength{\tabcolsep}{3pt}
\begin{tabular}{p{0.58\columnwidth}r}
\toprule
Audit item & Result \\
\midrule
Worker packet decisions & 41,452 \\
Packets processed at two or more nodes & 17,452 \\
Actual representation changes & 24,006 \\
Incompatible decisions & 0 \\
Exact-egress generation mismatches & 0 \\
Relay image-reconstruction events & 0 \\
Receiver format or placement errors & 0 \\
Unreconciled detour events & 0 \\
\bottomrule
\end{tabular}
\end{table}

\subsection{Paired Network and Task Outcomes}

In the sufficient profile, the two-hop service varies between 0.5 and
2.0~Mbit/s.  Both methods deliver all 600 coordinates of every image by the
12-second decision time, and neither records a congestion drop.  DINA reduces
receiver-side payload by 86.18\% (2.55 versus 18.43~MB) but cannot increase
coverage beyond 100\%.  Its 92.5\% accuracy is close to forwarding-only's
90.0\%; the paired directions are three DINA-only correct and two
forwarding-only correct decisions (McNemar $p=1$).  Thus the sufficient point
shows payload reduction without a significant task advantage.

The continuously fluctuating intermediate trace repeats 900, 600, 100, 40,
20, 20, 30, 60, 150, and 700 kbit/s epochs.  Object release is aligned to the
same trace phase in both methods.  DINA raises mean deadline coverage from
40.42\% to 72.60\% while reducing receiver-side payload from 7.45 MB to
0.79 MB.  It delivers more coordinates for every one of the 40 paired images,
with a mean gain of 193.075 tiles (exact paired sign test,
$p=1.82\times10^{-12}$).  The network change translates into a receiver-task
change: accuracy increases from 77.5\% to 95.0\%, fire recall from 80\% to
100\%, and false-fire rate decreases from 25\% to 10\%.  DINA alone is correct
on seven paired images, while forwarding-only alone is correct on none
(exact McNemar $p=0.0156$).

The extreme profile instead fluctuates between 5 and 37.5~kbit/s.  DINA still
raises mean deadline coverage from 4.14\% to 22.30\% and delivers more
coordinates for all 40 paired images (mean gain 108.975 tiles, exact sign-test
$p=1.82\times10^{-12}$).  This network gain is not sufficient for reliable
fire evidence: fire recall is only 15\% with DINA and 30\% with forwarding-only,
and the paired correctness directions do not establish an advantage (three
DINA-only versus six forwarding-only, McNemar $p=0.508$).  The corresponding
57.5\% and 65.0\% accuracies are dominated by all 20 non-fire images being
classified correctly.  This profile therefore lies outside the useful task
region despite its coverage gain.

\begin{table}[H]
\centering
\caption{Main paired network and receiver-task results.  Drops are fabric
egress-congestion drops over the complete paired run.}
\label{tab:main-results}
\scriptsize
\setlength{\tabcolsep}{2pt}
\begin{tabularx}{\linewidth}{>{\raggedright\arraybackslash}X
>{\raggedright\arraybackslash}Xrrrrrr}
\toprule
Region & Method & Coverage & \shortstack{Delivered\\bytes} & Drops & Accuracy &
\shortstack{Fire\\recall} & \shortstack{False-fire\\rate} \\
\midrule
Sufficient & Forwarding-only & 1.0000 & 18,432,000 & 0 & 0.900 & 1.00 & 0.20 \\
           & DINA            & 1.0000 & 2,547,356 & 0 & 0.925 & 0.95 & 0.10 \\
Intermediate & Forwarding-only & 0.4042 & 7,450,368 & 14,380 & 0.775 & 0.80 & 0.25 \\
             & DINA            & 0.7260 & 788,552 & 6,758 & 0.950 & 1.00 & 0.10 \\
Extreme & Forwarding-only & 0.0414 & 763,392 & 22,522 & 0.650 & 0.30 & 0.00 \\
        & DINA            & 0.2230 & 10,706 & 17,967 & 0.575 & 0.15 & 0.00 \\
\bottomrule
\end{tabularx}
\end{table}

\subsection{Packet Response within a Fluctuating Image Transfer}

\begin{figure}[H]
\centering
\includegraphics[width=\linewidth]{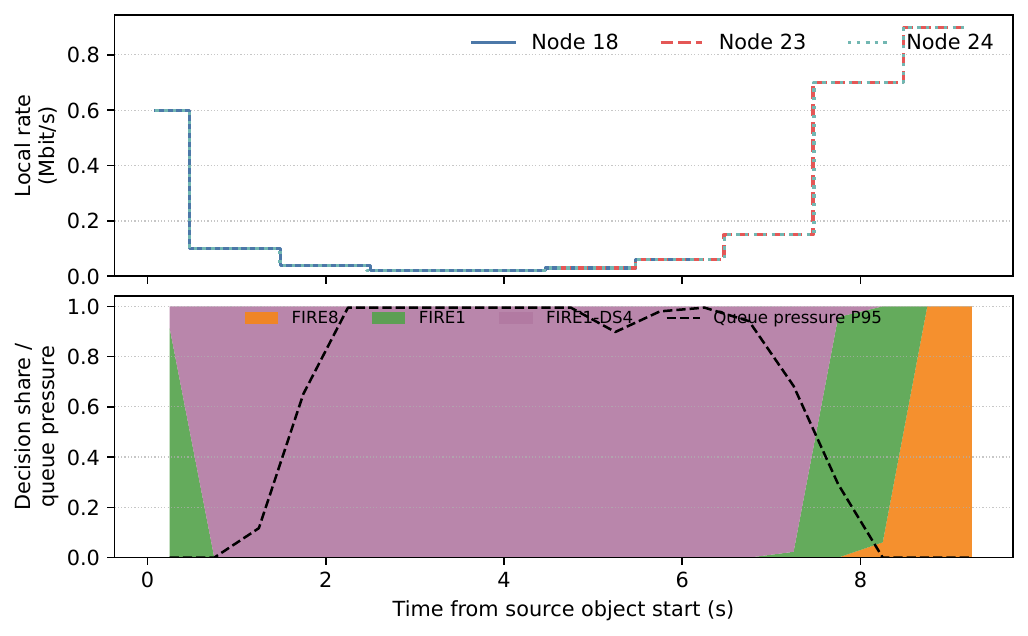}
\caption{Packet-local response during one image transfer in the fluctuating
intermediate profile.  The upper panel joins each on-path decision to its
exact-egress rate.  The lower panel bins selected representations and P95
queue pressure every 0.5 seconds.}
\label{fig:dynamic-response}
\end{figure}

Figure~\ref{fig:dynamic-response} follows one held-out image through 9.16
seconds of the intermediate profile.  The joined trace contains 1,042
decisions at nodes 18, 23, and 24 for 600 source packets.  Their observed
egress rates fall from 0.6~Mbit/s through 0.1, 0.04, and 0.02~Mbit/s before
recovering to 0.9~Mbit/s.  Low service and high queue pressure produce mostly
FIRE1-DS4 decisions; recovery produces later FIRE1 and
FIRE8 decisions.  The complete object trace contains 766
FIRE1-DS4, 184 FIRE1, and 92 FIRE8 node decisions.

\subsection{Cross-Task Reuse on RescueNet}

The RescueNet matrix changes both the task-related representations and the
receiver from binary fire classification to 11-class post-disaster semantic
segmentation.  Each of the 24 method--region runs uses the same 40 test images,
600 source tiles per image, phase alignment, queue, source-relative deadline,
network image, and receiver checkpoint.  The generated primary route is
$24\!\rightarrow\!4\!\rightarrow\!3\!\rightarrow\!7\!\rightarrow\!17$;
the installed backup is
$24\!\rightarrow\!14\!\rightarrow\!13\!\rightarrow\!18\!\rightarrow\!17$.
No route replacement occurs during a run, and every per-run and cross-run
contract check passes.

Table~\ref{tab:rescuenet-network} gives the DINA and forwarding results plus
the fixed representation with the highest observed dataset-level mIoU in each
region; all four fixed controls remain in the underlying comparison matrix.
At the sufficient point, DINA retains RGB24 and exactly matches forwarding in
coverage and mIoU.  Gray8, Gray4, and Edge1 all deliver every tile but reduce
mIoU, so unnecessary compaction is directly visible.

\begin{table}[H]
\centering
\caption{Frozen RescueNet network results.  The fixed column reports the
highest-mIoU fixed representation in each region.}
\label{tab:rescuenet-network}
\scriptsize
\setlength{\tabcolsep}{2pt}
\begin{tabularx}{\linewidth}{>{\raggedright\arraybackslash}Xrrrr
>{\raggedright\arraybackslash}Xrr}
\toprule
& \multicolumn{2}{c}{DINA} & \multicolumn{2}{c}{Forward only}
& & \multicolumn{2}{c}{Highest-mIoU fixed} \\
\cmidrule(lr){2-3}\cmidrule(lr){4-5}\cmidrule(lr){7-8}
Region & Coverage & mIoU & Coverage & mIoU & Representation & Coverage & mIoU \\
\midrule
Sufficient   & 1.0000 & 0.5391 & 1.0000 & 0.5391 & RGB24 & 1.0000 & 0.5391 \\
Intermediate & 0.7938 & 0.4846 & 0.3110 & 0.3738 & Gray4 & 0.8158 & 0.5053 \\
Composite    & 0.9165 & 0.5405 & 0.6561 & 0.4859 & Gray8 & 0.8170 & 0.5312 \\
Extreme      & 0.0415 & 0.0080 & 0.0291 & 0.0078 & Gray8 & 0.0373 & 0.0077 \\
\bottomrule
\end{tabularx}
\end{table}

The static intermediate point isolates a strong in-network processing gain
over unchanged forwarding.  DINA raises mean deadline coverage by 0.4827 with
a paired 95\% interval of $[0.4651,0.5007]$.  Across the 39 images with a
defined pair, the mean per-image foreground-mIoU difference is 0.1605 with
interval $[0.0978,0.2315]$; DINA is higher on 35 and lower on four.  Fixed
Gray4 nevertheless exceeds DINA at this one static point in both aggregate
coverage and aggregate mIoU.  The composite trace below evaluates the
selector's intended role under changing conditions.

The composite trace tests the reason for online selection.  DINA produces 36
naturally mixed canvases among 40 images and raises coverage from 0.6561 to
0.9165 relative to forwarding.  Its dataset-level foreground mIoU is 0.5405,
compared with 0.4859 for forwarding.  The paired mean differences are 0.2604
in coverage, with interval $[0.2117,0.3095]$, and 0.1089 in per-image mIoU,
with interval $[0.0547,0.1747]$.  DINA also has a higher aggregate mIoU than
every fixed representation.  Against Gray8, the highest-mIoU fixed control,
the paired per-image mean difference is 0.0166 with interval
$[-0.0043,0.0374]$, leaving the task advantage over this fixed control
statistically inconclusive.  Edge1 reaches 0.9985
coverage but only 0.3542 mIoU, demonstrating that minimizing bytes or
maximizing tile count alone is not the task objective.

\begin{figure}[H]
\centering
\includegraphics[width=\linewidth]{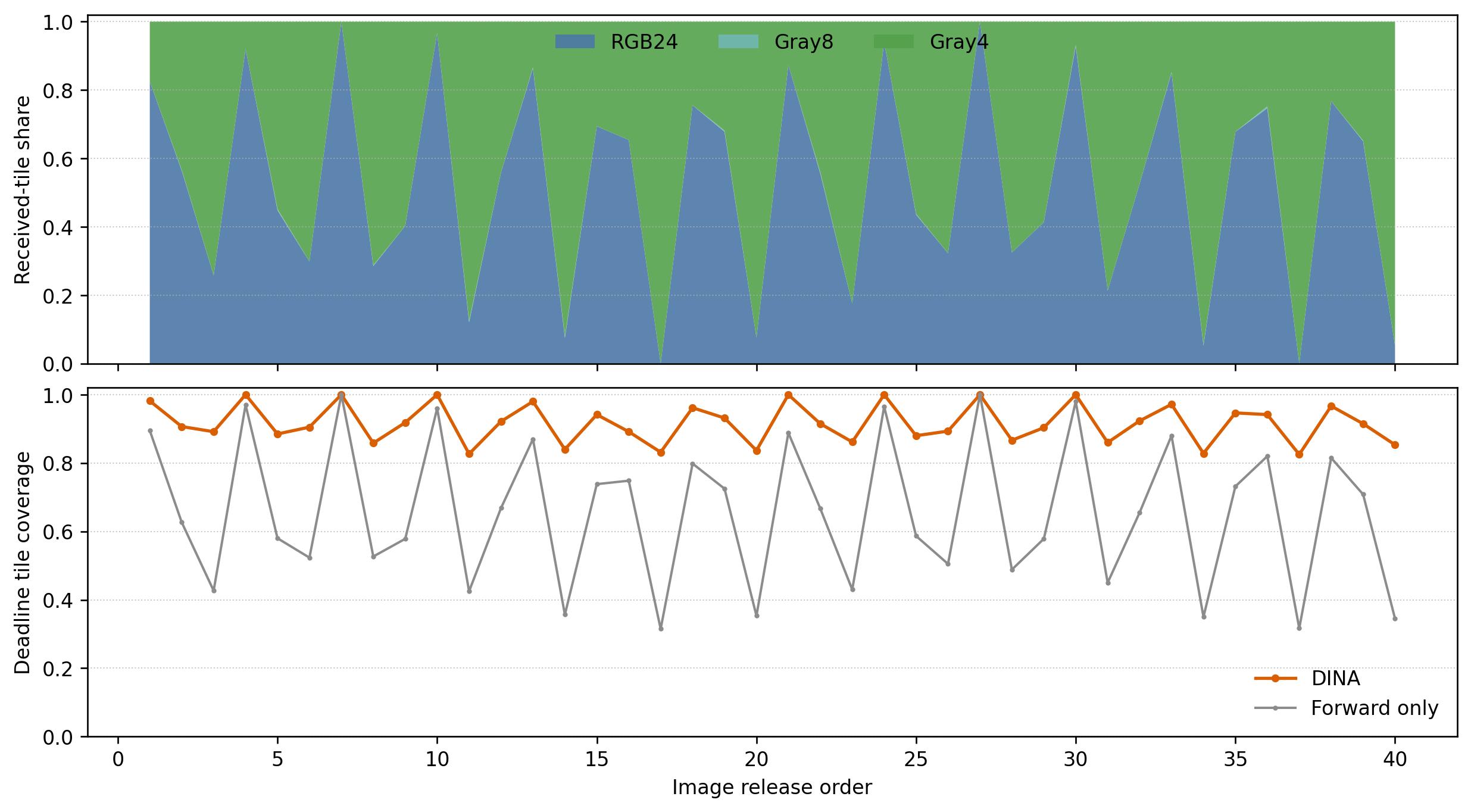}
\caption{Per-image response in the RescueNet composite trace.  The upper panel
shows the representation shares among DINA tiles received by the deadline;
the lower panel shows paired deadline coverage for DINA and forwarding only.
Each point uses the same frozen image order and trace phase.  The PNG is
generated by Python from per-image formal evidence.}
\label{fig:rescuenet-response}
\end{figure}

Figure~\ref{fig:rescuenet-response} shows how the composite aggregate arises.
The received DINA tiles alternate between retained RGB24 and compact Gray4 as
the phase-aligned service process changes; Gray8 appears only briefly.  The
lower panel pairs each image with forwarding under the identical release
phase.  Decisions occur at four capable nodes on the primary path, and later
nodes consume only the current representation carried by each packet.  The
response is consequently the realized sequence of independent node-local
choices.

At the extreme point, every method yields a dataset-level foreground mIoU
between 0.0064 and 0.0080.  Edge1 raises coverage only to 0.0456, and no paired
task comparison establishes a useful advantage.  This defines the common
physical and semantic failure boundary for the frozen packet rate, deadline,
queue, registry, and receiver.

\subsection{Online Processing Cost}

Table~\ref{tab:processing-cost} summarizes all 41,452 worker decisions in the
continuous trace.  The median measured decision path is 3.129~$\mu$s and the
P95 is 6.935~$\mu$s.  Compatible transformations cost more than an unchanged
decision, with a P95 of 7.843~$\mu$s across the three output formats.  The
80.23 decisions/s value is the offered rate of this experiment, not a worker
saturation limit.  Across the same 30-second resource samples, the three
eligible node containers use 1.86 CPU percentage points and 7.3~MiB more on
average than forwarding-only.  The packet carries a fixed 64-byte SPP header,
including a one-byte current representation identifier and no operation
history.

\begin{table}[t]
\centering
\caption{Measured packet-path cost under the continuous workload.}
\label{tab:processing-cost}
\small
\setlength{\tabcolsep}{3pt}
\begin{tabular}{p{0.58\columnwidth}r}
\toprule
Metric & Measured value \\
\midrule
All decisions, median / P95 & 3.129 / 6.935 $\mu$s \\
Unchanged, median / P95 & 1.355 / 2.616 $\mu$s \\
Transformed, median / P95 & 3.982 / 7.843 $\mu$s \\
FIRE8/FIRE1/FIRE1-DS4 transform P95 & 5.431 / 5.968 / 8.340 $\mu$s \\
Observed decisions & 41,452 (80.23 s$^{-1}$) \\
Eligible-node mean CPU, DINA / forward & 6.20\% / 4.34\% \\
Eligible-node mean memory, DINA / forward & 110.9 / 103.6 MiB \\
SPP header / current representation field & 64 B / 1 B \\
Operation history / relay image buffer & 0 B / 0 B \\
\bottomrule
\end{tabular}
\end{table}

\section{Discussion}
\label{sec:discussion}

\subsection{When Node-Local Adaptation Helps}

DINA uses the co-location expressed by Eq.~\eqref{eq:opportunity}: a node has
an actionable observation of its egress while the affected packet remains
mutable before that egress.  In the target high-dynamic regime, an endpoint
observe--return--act loop cannot reliably track independent intermediate-egress
variation: the reported condition may change before source-adapted packets
reach that hop.  The on-path node instead samples the current local condition
and acts on the packet already present there.

Source adaptation remains effective when the relevant state is stable and the
source still owns the affected data.  Once a packet reaches a changing
intermediate hop, node-local transformation remains actionable before egress
service; a transform after the weak hop cannot undo prior queueing or loss.
Across multiple capable nodes, the current representation lets each node
continue compatible processing without coordinating an image action.

\subsection{Operating and Semantic Boundaries}

The measured regions expose the method's operating boundary.  With sufficient
capacity, unchanged forwarding already delivers the task evidence.  In the
intermediate region, compact representations change delivery and the receiver
task.  Under extreme capacity, even the smallest registered representation
leaves inadequate spatial evidence.  Semantic correctness is the outcome of
the fixed receiver task on the realized canvas, reported together with the
network delivery that produced it.

The forest-fire color operators and RescueNet gray/edge operators retain
different task evidence.  Both use the same typed packet, local observation,
compatibility filtering, and on-path execution, while each application
supplies its registry, training material, and receiver.  The RescueNet static
intermediate result also shows that a fixed compact representation can exceed
the learned selector at one stable operating point; the selector's measured
role is adaptation across changing conditions.

\subsection{Scope and Deployment}

The continuous independent-image stream exercises offered load, queueing,
loss, route and link transitions, and mixed receiver canvases without adding
inter-frame codec dependencies.  Video deployment would require a codec-aware
packet contract exposing independently processable objects.  Other machine
tasks similarly provide their own packetization, registry, receiver, and
held-out task validation.

Payload transformation operates within a trusted domain shared by the
application and network.
The host publishes link state through a read-only table, the application
authorizes a finite packet format and operator set, and opaque encrypted
payloads require an explicitly transformable region.  The controlled 24-node
environment tests dynamic links, finite queues, routes, and containerized
workers reproducibly; field deployment additionally requires radio-state,
energy, security, and operational validation.

\section{Conclusion}
\label{sec:conclusion}

DINA turns an on-path node's local observation into an immediate action on the
packet already present at that node.  Self-describing spatial packets and
typed operator compatibility allow each packet to be processed independently
and to continue evolving across a multi-hop route.  The online node limits its
work to a frozen lightweight selector and a registered packet transform; full
task inference remains at the destination.  The destination measures semantic
success through its fixed machine task on the
coordinate-reconstructed canvas.  This design provides a concrete
in-network-computing mechanism for studying where local representation changes
expand the usable operating region of unstable mobile emergency networks.  A
forest-fire classifier and a RescueNet post-disaster segmenter show this
packet-path capability under two task-specific registries, while sufficient
and extreme profiles delimit where local adaptation provides no additional
task benefit or cannot overcome the physical channel.

\section*{Author Contributions}
Conceptualization, Z.R. and W.C.; methodology, Z.R.;
software, Z.R.; validation, Z.R. and T.Z.; formal analysis, Z.R. and T.Z.;
investigation, Z.R.; data curation, Z.R.; visualization, Z.R.; writing---original
draft preparation, Z.R.; writing---review and editing, Z.R., T.Z. and W.C.;
supervision, W.C.; project administration, W.C. All authors have read and
approved this manuscript.

\section*{Funding}
This research was funded by the National Key Research and Development
Program of China, grant number 2023YFC3011502.

\section*{Institutional Review Board Statement}
Not applicable.

\section*{Informed Consent Statement}
Not applicable.

\section*{Data Availability Statement}
The data and code supporting the findings of this study are
available from the corresponding author upon reasonable request.

\section*{Acknowledgments}
The background illustration in
Figure~\ref{fig:motivating-scenario} was generated using OpenAI's
image-generation system from an author-directed prompt.

\section*{Conflicts of Interest}
The authors declare no conflicts of interest. The funder
had no role in the design of the study; in the collection, analyses, or
interpretation of data; in the writing of the manuscript; or in the decision
to publish the results.

\bibliographystyle{unsrt}
\bibliography{references}

\end{document}